\documentclass[doublecol]{epl2}
\usepackage{amsfonts}

\title{Quantum ratchet control - harvesting on Landau-Zener transitions }
\author{L. Morales-Molina \inst{1}, S. Flach\inst{2}, J.B. Gong \inst{1,3}}
\institute{
  \inst{1}  Department of Physics and Center for Computational
Science and Engineering, National University of Singapore,
117542, Republic of Singapore \\
\inst{2}  Max-Planck-Institut f\"ur Physik Komplexer Systeme,
N\"othnitzer Str. 38, 01187 Dresden, Germany\\
\inst{3} NUS Graduate School for Integrative Sciences and
Engineering, Singapore
 117597, Republic of Singapore}


\abstract{
We control the current of a single particle quantum ratchet
by designing ramping schemes for experimentally accessible control
parameters. We harvest on
Landau-Zener transitions between Floquet states. Adiabatic and diabatic ramping
allow to control the resulting directed transport.
We find strong changes of the current
in  the adiabatic regime. Simple loops in control parameter space
with alternating adiabatic and diabatic ramping are proposed.
We obtain also current reversal.
Full desymmetrization of the quantum ratchet
increases the critical ramping speed which separates the adiabatic
from the diabatic regime.
}

\begin{document}

\maketitle
The ratchet effect concerns the rectification of transport in
the presence of zero mean forces. It was proposed to
 explain e.g. the molecular  motility in biological systems, and has been generalized to
 various other areas in physics
 \cite{Reim,hanggi}. Applications of ratchets range from
 biological systems \cite{julicher} to atomic physics \cite{ren1}.
 Especially in the case of light-matter interaction and laser-induced symmetry
 breaking with applications to cold atoms \cite{ren1}, control parameters are
 experimentally well accessible.

A symmetry analysis of Hamiltonian and dissipative classical single
particle ratchets was performed in Refs. \cite{Flach1,Flach2}.
A general sum rule to calculate the average velocity of classical ratchet
transport was obtained \cite{sumrule}.  Specific
dynamical mechanisms were also proposed \cite{Den}. Experiments 
on ratchet transport have also been carried
out using thermal cold rubidium and cesium atoms in optical
lattices \cite{ren1}. The time-dependent forces are applied to the
atoms by phase modulating the laser beams which form the optical
lattice \cite{ren1}. The dissipationless case can be easily
approached by using laser beams which generate far detuned standing
waves. In the latter case quantum features become
important. To address the quantum ratchet dynamics different models
have been proposed mostly
 based on kicked systems \cite{sumrule,tania,gongjb}.

In two recent studies of the quantum ratchet transport
\cite{EPL,PRA}, it was found that a directed current appears due to
a desymmetrization of the Floquet states of the system. A main
quantum feature of the rectification process is the resonant
interaction between Floquet states close to avoided crossings, which
leads to a strong enhancement of directed transport in a narrow
interval of the control parameters. In a further extension to these
studies, interaction between atoms has been considered as well
\cite{njp}.   In a different context,
 non-symmetric tunneling between bands in an asymmetric optical lattice
 has been reported
\cite{weitz}. The authors show that breaking the spatial symmetry induces changes
of the band gap leading consequently to a modification of the tunneling rate between bands.

Recent studies have also shown that it is possible to control the states of the energy
spectrum of electrons in semiconductor nanostructures by ramping the external
electrical field \cite{tambor}.
The authors  exploit the diabatic and adiabatic Landau-Zener transitions in
avoided crossings.

We study the effect of Landau-Zener tunneling transitions between
resonant Floquet states under non-symmetric ac forces for a quantum
ratchet. These transitions are controlled by the ramping speed of a
control parameter, which is related to the gap of the avoided
crossing of Floquet states. We find that directed transport is
significantly enhanced by adiabatically ramping parameters through
avoided crossings between resonant Floquet states, even when finally
ramping the systems far away from the resonance. We obtain current
reversal, and ways of increasing the critical ramping velocity which
separates adiabatic from diabatic ramping.  The results
constitute an extension of the control approach in \cite{tambor},
from navigating between eigenstates to navigating between Floquet
states.

Consider a dilute gas of atoms.  We can then, in good approximation, neglect
the interaction between particles.
The quantum dynamics for a particle moving in a periodic potential under the
influence of an ac force is described by the dimensionless
Schr\"odinger equation \cite{EPL,PRA}
 \begin{equation}
i \hbar \frac{\partial \psi}{\partial t}
=H \psi,  \label{shrogauge}
\end{equation}
where $H$ is the dimensionless one-particle Hamiltonian
\begin{equation}\label{hamilto0}
 H =\frac12\hat p^2 +v_{0}\cos(x)-xE(t)
\end{equation}
with $\hat p=-i\hbar \frac{\textstyle \partial}{\textstyle \partial
x}$, and $v_{0}=1$ being the potential depth. $\hbar$ is a
dimensionless effective Planck constant. The dimensionless ac field
satisfies $E(t+T)=E(t)$ with period $T$. As in \cite{EPL,PRA}, we
consider $E(t)=E_{1} \cos[\omega (t-t_{0})]+E_{2}\cos[2\omega
(t-t_{0})+\theta]$ with $t_{0}$ as initial time. By applying a
convenient gauge transformation \cite{gauge,EPL,PRA}, we get

\begin{equation}\label{hamilto}
 H=\frac12[\hat p-A(t)]^2 +\cos(x),
\end{equation}
where
 $A(t)=-E_{1} \sin[\omega
(t-t_{0})]/\omega-E_{2} \sin[2\omega (t-t_{0})+\theta]/2 \omega$ is the vector
potential \cite{EPL,PRA}.

\begin{figure*}
\onefigure[width=13cm,height=8cm]{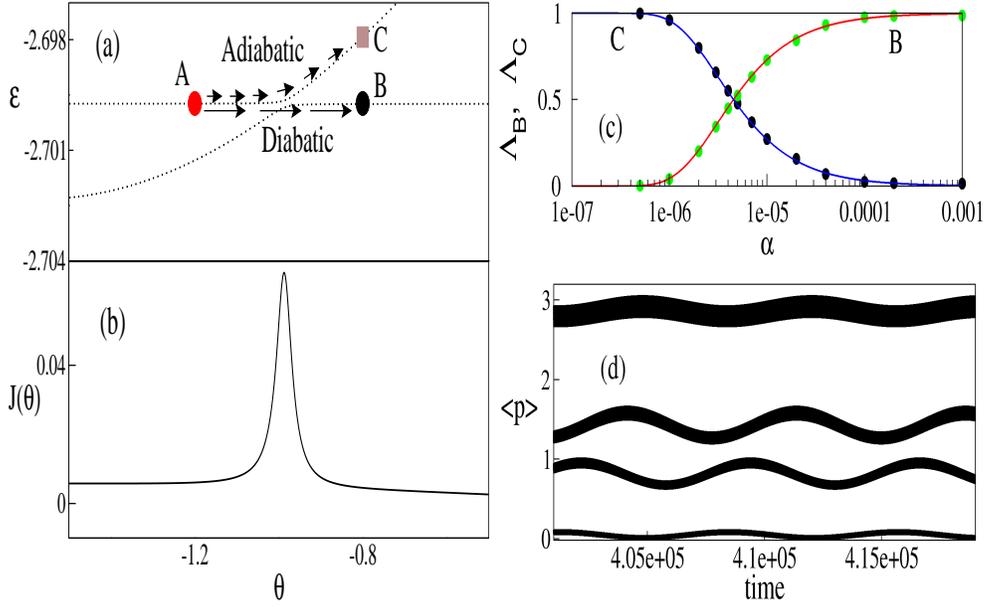}
\hspace{1.5cm}
\caption{ (a) Avoided crossings between two Floquet states which lead to an
  enhancement of the current (no ramping). The arrows indicate the
  direction of the ramping with initial state $A$. (b) Peak of the current due to the resonant states
  in the above avoided crossing.
(c): Transition probability $\Lambda$ vs  $\alpha$ (logarithmic scale) in the basis
of states $B$ and $C$ (see text for details). Circles: Numerical
computation of Eq.(\ref{Eq:transition}). The solid lines are the
fitting of Eq.(\ref{Eq:probability}) to the circles:  $P$ (red
line),  $1-P$ (blue line).(d): Momentum evolution of the final
state. The ramp is performed  from $\theta=-1.2$ to $\theta=-0.8$.
Momenta from top to bottom correspond to $\alpha=10^{-6}, 5\times
10^{-6}, 10^{-5},10^{-3}$, respectively. $t_{ini}=0$.
 The parameters are $E_{1}=3.26$, $E_{1}=1.2$ and $\omega=3$.}
\label{Fig:fullbif1}
\end{figure*}

 The Hamiltonian in Eq.(\ref{hamilto0}) is a periodic function of
time. Hence the solutions of Eq.(\ref{shrogauge}),
 $|\psi(t+t_0)\rangle = U(t_{0},t+t_0)|\psi(t_0)\rangle$, can be characterized by
 the eigenfunctions of $U(t_0,T+t_0)$,  which satisfy the relation:
\begin{equation}\label{eq:floquet}
 |\psi_{\beta}(t)\rangle=
e^{-i\frac{ \epsilon_{\beta}}{ T} t} |\phi_{\beta}(t)\rangle, \,\,\,\
|\phi_{\beta}(t+T)\rangle=|\phi_{\beta}(t)\rangle.
\end{equation}

The quasienergies $\epsilon_{\beta}$ $(-\pi < \epsilon_{\beta} <
\pi)$ and the Floquet eigenstates can be obtained as solutions of
the eigenvalue problem of the Floquet operator $
U(0,T)|\psi_{\beta}(0)\rangle = e^{-i
\epsilon_{\beta}}|\psi_{\beta}(0)\rangle.$

Due to
Bloch's theorem, discrete translational invariance of Eq.(\ref{hamilto})
implies that Floquet states are characterized by a
 quasimomentum $\hbar \kappa$ with
$ |\psi_{\beta}(x+L)\rangle = {\rm e}^{i 2 \pi \kappa}|
\psi_{\beta}(x)\rangle$. We choose $\kappa=0$ (a treatment
excluding studies of evolution of wavepackets)  because our
interest is in how a nonzero flux of atoms emerge from an initial
state with almost zero momentum. The $\kappa=0$ initial state can be
produced by, for instance, the ground state of an elongated
(shallow) trap potential of a condensate.  This consideration allows
us to use periodic boundary conditions for Eq. (\ref{shrogauge}),
with spatial period $L=2 \pi$. Hence the
 wave function can be expanded in the plane wave
 eigenbasis of the momentum operator
$\hat{p}$, $|n \rangle=\frac{1}{\sqrt{2 \pi}} e^{i n x}$ \cite{PRA}.
The Floquet operator is obtained by solving
Eqs.(\ref{shrogauge}-\ref{hamilto}). Computational details
are given in  \cite{PRA}.

Breaking the time-reversal and shift  symmetries of the Floquet operator \cite{Graham}
generates a nonzero average
momentum of the Floquet states, and results in a nonzero directed transport \cite{EPL,PRA}.

The asymptotic current of the system for $t \gg T$ is obtained through the expression
$J(t_0)=\sum_{\beta} \langle p \rangle_{\beta}
|C_{\beta}(t_{0})|^{2}$
 \cite{PRA}, where $\langle p \rangle_{\beta}$ are the Floquet state
 momenta and $C_{\beta}(t_{0})$ are the expansion
 coefficients of the initial wave function in the basis of Floquet states.
 The current is a function of the relative phase $\theta$ and the initial
 time $t_{0}$, i.e. $J(t_0,\theta)$. After averaging over the
 initial time the current becomes a function of $\theta$ only and fulfills the
 relation $J(\theta)=-J(\theta+\pi) = -J(-\theta) $ \cite{EPL,PRA}.

 \begin{figure*}
\begin{center}
\onefigure[width=12cm,height=6cm]{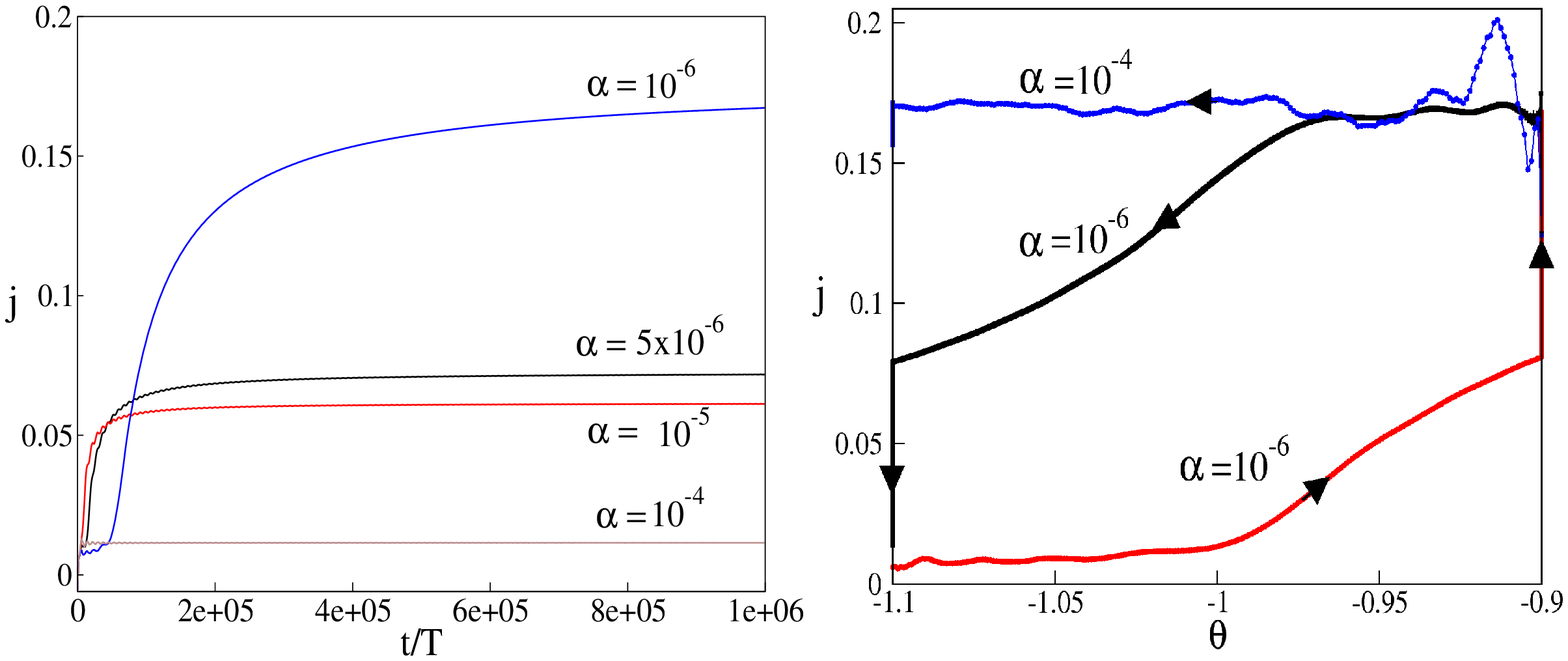}
\caption{Left panel: Current of the state $|0\rangle$ as a function of
  time in units of the period $T$.
 Ramping is performed from $\theta=-1.1$ till $\theta=-0.9$. $t_{ini}=1000$.
Right panel: Current $j$ vs $\theta$ as we ramp forward and backward with
different rates.
We ramp forward from  $\theta=-1.1$ till $\theta=-0.9$ with $\alpha=10^{-6}$.
 Then we ramp from $\theta=-0.9$ to  $\theta=-1.1$ with
$\alpha= 10^{-4}, 10^{-6}$. The arrows indicate the tendency of the current
value as time evolves. In both panels $t_{0}=0$.
 Parameters are the same as in
Fig.\ref{Fig:fullbif1}.}\label{Fig:current}
\end{center}
\end{figure*}


For a given value of the control parameter $\theta$, the initial state
$|n=0\rangle$ will most strongly overlap with e.g. a certain Floquet state, yielding
a nonzero asymptotic current. Variation of $\theta$ will lead to avoided crossings
between quasienergies, i.e. to a resonant interaction between pairs of Floquet states, and are confined to
a small interval in $\theta$. Inside the avoided crossing region the Floquet
states hybridize. If the second (new) Floquet state has a large mean momentum,
the asymptotic current will resonantly increase. But upon leaving the avoided crossing
region, the Floquet states cease to interact and to hybridize, and the asymptotic
current returns to its nonresonant (lower) value. We will therefore ramp
the control parameter $\theta$ in time. Depending on the ramping velocity we expect
to either reobserve the results for the case without ramping (diabatic Landau-Zener regime),
or to slowly populate the second state (adiabatic regime). In the second case, after
leaving the avoided crossing region, the quantum state will reside mainly in the second
Floquet state and keep the large asymptotic current.

Hereafter the analysis is done for $\hbar=0.2$.
Fig.\ref{Fig:fullbif1} displays an avoided crossing of Floquet
states that leads to an enhancement of the current. The resonance
takes place between a state $A$ located in the chaotic layer of the
corresponding classical model, and a transporting state $C$  (see
\cite{EPL,PRA} for details). We first consider an initial quantum
state which exactly equals the Floquet state $A$ in
Fig.\ref{Fig:fullbif1}. Then we ramp $\theta$ linearly in time, viz.

\begin{eqnarray}
\label{Eq:rampafortaleza}
 & \theta(t)= \Lambda(\theta_{1},\theta_{2},\alpha,t) \hspace{5cm} \nonumber \\
&\equiv\left\{ \begin{array}{c}
~~\theta_{1},  \hspace {2cm} ~ t \le t_{ini} ~~~~~~~~~ \\
 \theta_{1}+\alpha (t-t_{ini}), \hspace {0.3cm} t_{ini} < t \le
 t_{ini}+(\theta_{2}-\theta_{1})/\alpha ~~\\
~~ \theta_{2} , \hspace {0.5cm} ~~~~~~~~   t>
t_{ini}+(\theta_{2}-\theta_{1})/\alpha,
\end{array}\right.
\end{eqnarray}
where $\alpha$ is the ramping rate and $t_{ini}$  is the time at
which the ramping process starts. $\theta_{1}$ and $\theta_{2}$ are
the initial and final values.

In the adiabatic regime $\alpha \rightarrow 0$,
state A  transforms into the transporting state $C$. The ramping time
being finite, we
estimate the deviation
from the strict adiabatic behavior. We define
\begin{equation}\label{Eq:span}
\psi_{f}=a(t)\psi_{C}+b(t) \psi_{B}
\end{equation}
as the final state in the basis of states $C$ and $B$.
Next we define
\cite{breuer}
\begin{equation}\label{Eq:transition}
 \Lambda_{C,B}=\frac{1}{T}\int_{0}^{T} dt |\langle
 \psi_{f}|\psi_{C,B}\rangle |^2,
\end{equation}
as the transition probability of the initial state $A$
into the states $B$ and
$C$, after ramping $\theta$ (Fig.\ref{Fig:fullbif1}).

For adiabatic ramping, we assume that variations in $\theta$ are
negligible during one period of the force, which allows us to separate
the time scales of ac driving and ramping \cite{breuer}. Then, we can use the standard
 Landau-Zener approach \cite{landau} and derive the transition
 probability,  from which we estimate the ramping rate needed for an adiabatic
 process.
The transition probability for Floquet states is given by \cite{breuer,landauholtaus},
\begin{equation}\label{Eq:probability}
P=\exp\left[-\frac{\pi\,\, \delta \epsilon\,\,\delta \theta}{2 \hbar\,\ \alpha}\right],
\end{equation}
where  $\delta \theta$ is the $\theta$- interval in which the
quasienergy difference increases by a factor of $\sqrt{2}$
\cite{breuer}. $\delta \epsilon$ denotes the quasienergy gap. Thus,
we find $\delta \epsilon \approx 2\times  10^{-4}$, $\delta \theta
\sim 0.01$ for the avoided crossing in Fig.\ref{Fig:fullbif1}. This
leads to the rough estimate $\frac{\textstyle \delta \epsilon \delta
\theta}{\textstyle 2 \hbar}=5 \times 10^{-6}$. To be more accurate
we fit the transition probabilities in Fig.\ref{Fig:fullbif1} using
Eq.(8) and obtain $\frac{\textstyle \delta \epsilon \delta
\theta}{\textstyle 2 \hbar}\approx 10^{-6}$. Therefore, for $\alpha
< 10^{-6}$, the adiabatic regime is realized. Indeed, from
Fig.\ref{Fig:fullbif1}c, we see that, as the ramping rate decreases,
the transition probability to the state $C$ increases consequently.
A crossover takes place at $\alpha\approx 5 \times 10^{-6}$ as the
system reaches one half of the transition probability.

The final state is a superposition of Floquet states $B$ and $C$
(see (\ref{Eq:span})), and therefore displays interference effects
\cite{breuer}. A signature of that interference are the oscillations
experienced by the momentum of the final state
Fig.\ref{Fig:fullbif1}d. These oscillations take place around a
nonzero value, which increases as the system approaches to the
adiabatic behavior. Using Eqs.(\ref{Eq:span}) and
(\ref{eq:floquet}), it is  straightforward to find
\begin{eqnarray}\label{Eq:momfina}
 \bar p&=|a|^2 \bar p_{C} + |b|^2 \bar p_{B}+a^{*} b
 \exp\left(-i\frac{\epsilon_{B}-\epsilon_{C}}{T}t\right) \langle \phi_{C}
 |\hat{p}| \phi_{B}\rangle \nonumber \\
  &+ a b^{*}
 \exp\left(-i\frac{\epsilon_{C}-\epsilon_{B}}{T}t\right) \langle \phi_{B}|
 \hat{p}| \phi_{C}\rangle  \; .
\end{eqnarray}
The last two terms in Eq.(\ref{Eq:momfina}) generate oscillations,
which are slow compared to the period of the driving force.
The period of the
slow oscillations in Fig.\ref{Fig:fullbif1}d is in agreement with
the estimation $\frac{\textstyle 1}{\textstyle
  |\epsilon_{C}-\epsilon_{B}|}\sim \tau
\approx 10^4$. In the extreme adiabatic limit $b=0$, and consequently the last two
terms in Eq.(\ref{Eq:momfina}) disappear.

Let us now address the effect of adiabatic ramping on the
generation of directed transport.
To this end, we ramp $\theta$ within the interval $\theta \in(-1.1,-0.9)$,
where no other avoided crossings appear. We
consider the  initial state $|0\rangle$ with zero momentum. It
overlaps strongly with the state A depicted by a circle in
Fig.\ref{Fig:fullbif1}a.  By adiabatically ramping
through the resonant region, the quantum state will strongly overlap
with the transporting state $C$, thus leading to an increase of the current.

We compute the time-dependent current as
$j=\frac{\textstyle 1}{\textstyle (t -t_{0})} \displaystyle \int_{t_{0}}^{t}\bar p \!\ dt $ with
$\bar p =\langle \psi|\hat p|\psi \rangle$ which becomes the asymptotic
current $J$ in the  limit $t\rightarrow \infty$.
Fig.\ref{Fig:current} displays the current evolution as we ramp
with different rates. We find that the final current increases by an order
of magnitude when the ramping velocity is lowered from the diabatic into
the adiabatic regime. If we ramp again, but backwards, with the same adiabatic rate,
we will more or less return to the original state and momentum. However,
if we ramp fast (diabatically) backwards, we will perform a Landau-Zener transition,
and arrive at the initial value $\theta = -1.1$, but now with a much larger current,
than the one we started with!
Fig.\ref{Fig:current} shows the corresponding numerical results,
as we ramp forward and backward with different ramping velocities.
We ramp forward and backward between $\theta=-1.1$ and
$\theta=-0.9$. When ramping forward we use $\alpha=10^{-6}$ which corresponds to an
adiabatic regime. During the backward ramping we use two values of $\alpha$. Whereas in an adiabatic
backward ramp the current becomes small again, for the diabatic ramp the
current remains around its maximum value  [Fig.\ref{Fig:current}].

\begin{figure}
\begin{center}
\onefigure[width=7cm,height=12cm]{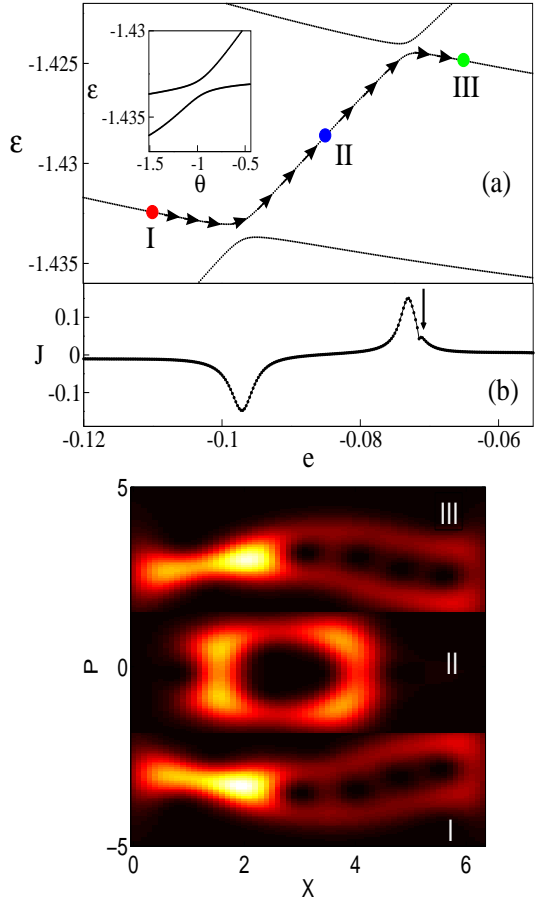} \caption{Upper
panel: (a) Quasienergy $\varepsilon$ vs amplitude $e$ of the second
  harmonic in the potential $v(x)$. The arrows indicate the ramping process
  starting in I, then passing through the intermediate state
  II, finally ending up  in III. Inset: Quasienergy vs $\theta$ at
  the resonance point $e=-0.097$. (b) Current $J$ vs $e$. The arrow
  indicates a narrow peak which appears due to an avoided crossing not shown
  in the graphic. Lower panel: Husimi functions of the states labeled by Roman
 numbers in upper panel (a). The intensity scales, in descending order, from yellow to black. We use the Husimi function defined in \cite{PRA}.
The parameters are $\theta=-0.99$ and
 $\theta_{p}=-\pi/2$. The other parameters are the same as in Fig.\ref{Fig:fullbif1}.}\label{Fig:currentramp}
\end{center}
\end{figure}

The gap of an avoided crossing determines the maximum adiabatic
ramping rate. In order to increase that rate, the avoided crossing
gap has to be increased.  At the same time,  for a Landau-Zener
description to be applicable, the gap value should remain
considerably smaller than the average level spacing. The gap value
is determined by the strength of interaction between Floquet states.
The interaction strength in turn is the larger, the more asymmetric
the Floquet operator is. Motivated by this observation as
well as the work of Ref. \cite{weitz},  we further consider a
bichromatic optical lattice potential to enhance the gap value. In
particular, we consider the potential $v(x)=\cos(x)+e
\cos(2x+\theta_{p})$ as in Ref. \cite{weitz}.  We
compute the quasienergy spectrum at $\theta=-0.99$ and ramp $e$.
Fig. \ref{Fig:currentramp} shows two avoided crossings of different
states which lead to peaks with opposite current signs. The
corresponding Husimi functions (see \cite{EPL,PRA} for details)
reveal that such peaks appear due to resonances between a state in
the chaotic layer and transporting states that carry opposite mean
momenta.

We use that structure and ramp adiabatically through this parameter
window. As a result the initial state I with nonzero momentum
transforms into a state with opposite momentum, and thus we reverse
the current  ($\alpha_{2}$ denotes the ramping rate for $e$).

We first ramp $e$ from $-0.11$ to $-0.085$ with state I as initial
input. After that, we compute the momentum of the obtained
intermediate state as a function of time. Then we ramp $e$ from
$-0.085$ to $-0.065$, with the intermediate state as initial input.
Then, we compute the evolution of the momentum of the final state.

\begin{figure}
\onefigure[width=6cm,height=12cm]{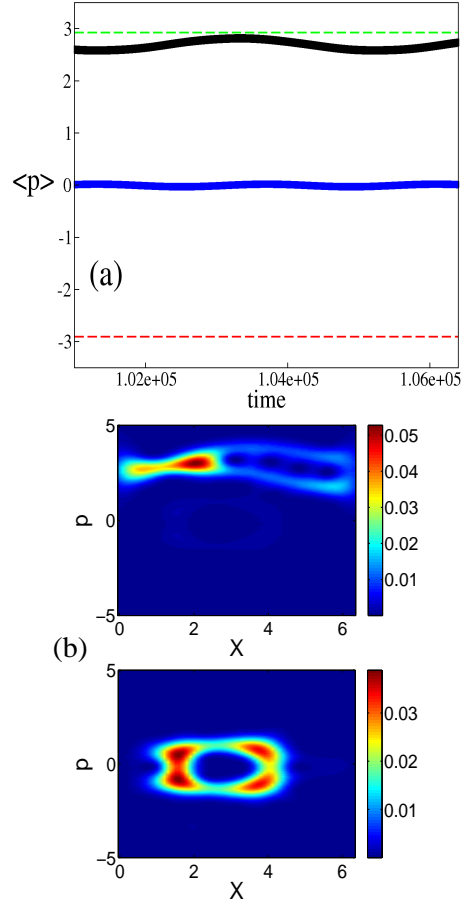}
\caption{(a) Momenta of the intermediate and final states while ramping $e$
  as a function of time depicted by thick curves. The thickness of the curves is a result of periodic oscillations of the momenta with period $T$, which are not resolved by the eye in this time scale.   Mean momentum for the states I, II and III
 in  Fig.\ref{Fig:currentramp} are depicted by dashed lines. The dashed line
 for the state II appears superimposed to the momentum of the intermediate state. The ramping rate is $\alpha_{2}=2.5\times 10^{-7}$. (b)  Husimi functions for intermediate and
 final states at $t=50998\,\ T$.
The other parameters are the same as in
Fig.\ref{Fig:currentramp}.}\label{Fig:transition2}
\end{figure}

Fig. \ref{Fig:transition2} shows the momenta of the intermediate and
final states as a function of time. Both show good agreement with
the momenta of the states II and III of Fig.\ref{Fig:currentramp},
respectively. The oscillations indicate interference due to the
coupling between the states. Husimi functions for states in
adiabatic ramping show a good correspondence to the Floquet states
II and III in Fig.\ref{Fig:currentramp}. We therefore manage to
reverse and increase the current value significantly.

If ramping $\theta$ instead within the interval $(-1.5,-0.5)$ (see
inset of Fig.\ref{Fig:currentramp}a), we find that, for an adiabatic
ramping, the transition rate should be below $\alpha=4 \times
10^{-5}$. This value is much larger than the one for the symmetric
case $e=0$. Certainly, further increase of the gap values can be
obtained by tuning other system parameters as well, thus further
reducing the time needed for the adiabatic ramping.

To summarize, we ramped control parameters through avoided crossings
of Floquet states with different average current characteristics.
When diabatically ramping, the crossing is barely visible due to
Landau-Zener transitions. For adiabatic ramping, the final quantum
state will strongly change its transport properties. One can
therefore switch the average current from small to large values or
vice versa. By first ramping adiabatically, and then ramping back
diabatically, we can therefore reach a completely different quantum
state at one and the same value of the control parameter. Harvesting
on several consecutive crossings, it is possible to achieve current
reversals. In addition, by considering a bichromatic optical
lattice potential,  we show it is possible to increase the ramping
speed while maintaining adiabaticity.

One of the authors (J.B. GONG) is supported by the start-up funding
(WBS grant No. R-144-050-193-101 and No. R-144-050-193-133) and the
NUS ``YIA'' funding (WBS grant No. R-144-000-195-123), both from the
National University of Singapore.


\begin{thebibliography}{1000}
\bibitem{Reim} \Name{Reimann P.}
        \REVIEW{Phys. Rep.} {361} {2002} {57}.
        \bibitem{hanggi} Astumian R. D. and Hanggi P., Physics Today
        {55}, No. 11, (2002) 33.

\bibitem{julicher}\Name{ J\"{u}licher F., Ajdari A., and Prost J.}
             \REVIEW{ Rev. Mod. Phys.} {69} {1997} {1269}.

\bibitem{ren1} \Name{Schiavoni M., Sanchez-Palencia L., Renzoni F., and GrynbergG.}
              \REVIEW{ Phys. Rev. Lett.} {90}{2003}{094101};
                  \Name{ Jones P. H.,  Goonasekera M. , and Renzoni F.}
               \REVIEW{ Phys. Rev. Lett.} {93}{2004} {073904};
               \Name{Gommers R., Bergamini S., and Renzoni F.}
                \REVIEW{ Phys. Rev. Lett.} {95} {2005} {073003};
              \Name {Gommers R.,  Denisov S., and Renzoni F.}
               \REVIEW{ Phys. Rev. Lett.} {96} {2006} {240604}.
\bibitem{Flach1} \Name{Flach S., Yevtushenko O., and Zolotaryuk Y.}
                 \REVIEW{Phys. Rev. Lett.} {84} {2000} {2358}.
\bibitem{Flach2} \Name{Yevtushenko O., Flach S., Zolotaryuk Y.,
                  and Ovchinnikov A. A.}
                 \REVIEW{ Europhys. Lett.}{54}{2001}{141}.
\bibitem{sumrule} \Name{Schanz H., Otto M.-F., Ketzmerick R., Dittrich T.}
                \REVIEW{Phys. Rev. Lett.} {87} {2001} {070601};
                \Name{Schanz H., Dittrich T., and Ketzmerick R.}
                \REVIEW{ Phys. Rev. E}{71}{2005}{026228}.

\bibitem{Den}\Name{Denisov S. \etal}
              \REVIEW{ Phys. Rev. E}{66}{2002}{041104}.
\bibitem{tania} \Name{Monteiro T. S., Dando P. A., Hutchings N. A. C., and
                Isherwood M. R.}
               \REVIEW{Phys. Rev. Lett.}{89}{2002}{194102}.
\bibitem{gongjb}
                \Name{Gong J. and Brumer P.}
                 \REVIEW{Phys. Rev. E} {70}{2004}{016202};
                 \Name{Gong J. and Brumer P.}
                 \REVIEW{Annu. Rev. Phys. Chem.} {56}{2005}{1};
                 \Name{Gong J. and Brumer P.}
                 \REVIEW{Phys. Rev. Lett.} {97}{2006}{240602}.

\bibitem{EPL} \Name{Denisov S., Morales-Molina L., and Flach S.}
              \REVIEW{Europhys. Lett.} {79} {2007} {10007}.

\bibitem{PRA} \Name{ Denisov S., Morales-Molina L., Flach S., H\"anggi P.}
                \REVIEW{ Phys. Rev. A} {75} {2007} {063424}.

\bibitem{njp} \Name{ Morales-Molina L. and Flach S.}
                \REVIEW{New Journal of Physics} {10} {2008} {013008}.


\bibitem{weitz} \Name{Salger T., Geckerler C., Kling S., and Weitz M.}
                 \REVIEW{ Phys. Rev. Lett.} {99} {2007} {190405}.

\bibitem{tambor} \Name{Wisniacki D. A., Murgida E. G., and  Tamborena P. I.}
                 \REVIEW{ Phys. Rev. Lett.} {99} {2007} {036806}.

\bibitem{gauge}\Name{Latka M., Grigolini P., and West B. J.}
               \REVIEW{  \ Phys.\ Rev. \ A}{50}{1994}{1071};
                \Name { Henseler M., Dittrich T., and Richter K.}
                \REVIEW{ Phys. Rev. E}{64} {2001}{046218}.

\bibitem{Graham} \Name{Graham R. and Keymer J.}
                \REVIEW{Phys. Rev. A} {44} {1991} {6281}.

\bibitem{breuer} \Name{ Breuer H. P. and Holthaus M.}
                  \REVIEW{Z. Phys. D} {11} {1989} {1}.

\bibitem{landau} \Name{Landau L. D.}
               \REVIEW{Phys. Z. Sowjetunion} {2} {1932} {46};
               \Name{Zener G.}
               \REVIEW{Proc. R. Soc. London, Ser. A} {137} {1932} {696}.

\bibitem{landauholtaus} \Name{Breuer H. and Holthaus M.}
                  \REVIEW{Phys. Lett. A} {140} {1989} {507}.


\end{thebibliography}
\end{document}